\documentclass[twocolumn,showpacs,preprintnumbers,amsmath,amssymb]{revtex4}
\usepackage{graphicx}
\usepackage{dcolumn}
\usepackage{bm}

\begin{document}

\title{Entanglement evolution of two qubits under noisy environments}
\author{Jun-Gang Li}
 \email{jungl@bit.edu.cn}
\author{Jian Zou}%
 \email{zoujian@bit.edu.cn}
 \author{Bin Shao}%
 \affiliation{Department of Physics, Beijing Institute of Technology, Beijing 100081, People's Republic of China}%
 \affiliation{Key Laboratory of Cluster Science of Ministry of Education, Beijing Institute of Technology, Beijing 100081, People's Republic of China}%
\date{\today}
\begin{abstract}
The entanglement evolution of two qubits under local, single- and
two- sided noisy channels is investigated. It is found that for all
pure initial states, the entanglement under a one-sided noisy
channel is completely determined by the maximal trace distance which
is the main element to construct the measure of non-Markovianity.
For the two-sided noisy channel case, when the qubits are initially
prepared in a general class of states, no matter pure or mixed, the
entanglement can be expressed as the products of initial
entanglement and the channels' action on the maximally entangled
state.
\end{abstract}

\maketitle

\section{Introduction}

In realistic quantum information processing \cite{Nielsen,Bennett},
quantum systems are open to essentially uncontrollable environments
that act as sources of decoherence and
dissipation\cite{Breuerb,Gardinerb}. Therefore the study of
entanglement evolution of quantum systems taking into account the
effect of environment has become more and more
interesting\cite{TYu,Dodd,Zczkowski,Konrad,Chang,Zong}. In
Ref.\cite{Konrad} Konrad \textit{et al}. provided a direct
relationship between the initial and final entanglement of two
qubits with one qubit subject to incoherent dynamics. It has been
shown that given any one-sided quantum channel, the concurrence of
output state corresponding to any initial pure input state of
interest can always be equivalently obtained by the product of the
concurrence of input state and that of the output state with the
maximally entangled state as an input state. However, for the
two-sided quantum channel case or the mixed initial state case, the
product of the two concurrences only provides an upper bound for the
concurrence of interest. It is obviously important to find the exact
relations of that for two-sided noisy channel case and that for the
mixed initial state case.

Over the past decades, the conventionally employed Markovian
approximation with the assumption of an infinitely short correlation
time of environment has experienced more and more challenges due to
the advance of experimental techniques\cite{Breuerb}. Recent studies
\cite{Piilo1,Breuer3,Maniscalco,Ferraro,Budini,Dijkstra,Vacchini,Daffer,Chou,Huang,Shabani,Chru,Khalili,Fanchini,Bellomo}
have shown that non-Markovian quantum processes play an increasingly
important role in many fields of physics. In order to quantitatively
study the non-Markovian dynamics, following the first computable
measure of ¡°Markovianity¡± for quantum channels introduced in
Ref.\cite{Wolf}, some measures for the degree of non-Markovianity
have been introduced\cite{BreuerNd,Rivas}. It is very interesting to
investigate the explicit relationship between the entanglement
evolution and the non-Markovianity.

In this paper, we consider the entanglement evolution of two qubits
under local noisy channels. The main aim of this work is to analyze
if and to what extent can the factorization law given in
Ref.\cite{Konrad} be generalized. As a central result, we
demonstrate that for the two-sided noisy channels, even when the
qubits are initially prepared in some mixed states, the concurrence
can be expressed as the products of the initial concurrence and the
channels' action on the maximally entangled state.

The paper is organized as follows: In Sec.\ref{model}, we present
the model and its analytical solution. In Sec.\ref{entdynamic} we
give the entanglement dynamics of two qubits under local, single-
and two- sided non-Markovian channels. In Sec.\ref{applications} two
applications are given. Finally we give a conclusion of our results
in Sec.\ref{conclusion}.

\section{The model}\label{model}
We consider a system formed by two noninteracting parts $A$ and $B$,
each part consisting of a qubit $s = a, b$ locally interacting
respectively with a reservoir $R_S = R_A,R_B$. Each qubit and the
corresponding reservoir are initially considered independent. The
dynamics of each part, consisting a qubit with excited state
$|e\rangle$ and ground state $|g\rangle$ which is coupled to a
reservoir of field modes initially in the vacuum state, can be
represented by the reduced density matrix\cite{Breuerb}
\begin{equation}\label{singlequbit}
\rho^{S}(t)=\left(
\begin{array}{cc}
\rho^S_{ee}(0)|h_S(t)|^2&\rho^S_{eg}(0)h_S(t)\\
\rho^S_{ge}(0)h^*_S(t)&1-\rho^S_{ee}(0)|h_S(t)|^2\\
\end{array}
\right)\\
\end{equation}
in the qubit basis $\{|e\rangle,|g\rangle\}$, where the superscript
$S=A,B$ represents part $S$. The function $h_S(t)$ ($h(t)$ for
short) is defined as the solution of the integrodifferential
equation
\begin{equation}\label{difc1}
\frac{d}{dt}h(t)=-\int_0^t dt_1f(t-t_1)h(t),
\end{equation}
with $f(t-t_1)$ denotes the two-point reservoir correlation function
which can be written as the Fourier Transform of the spectral
density $J(\omega)$:
\begin{equation}\label{ff}
f(t-t_1)=\int d\omega J(\omega)\exp[i(\omega_0-\omega)(t-t_1)].
\end{equation}
The exact form of $h(t)$ thus depends on the particular choice of
the spectral density of the reservoir. It should be noted that the
dynamical map of the model can be Markovian (iff $|h(t)|$ is a
monotonically decreasing function of time) and non-Markovian (iff
$|h(t)|$ increases at some times)\cite{Laine}.

Next, we are ready to use the reduced density matrix elements given
in Eq.(\ref{singlequbit}) to construct the reduced density matrix
for the two-qubit system.  Following the procedure presented in
Ref.\cite{Bellomo}, we find that in the standard product basis
$B=\{|1\rangle=|ee\rangle,
|2\rangle=|eg\rangle,|3\rangle=|ge\rangle,|4\rangle=|gg\rangle,\}$,
the diagonal elements of the reduced density matrix $\rho(t)$ for
the two-qubits system can be written as
\begin{equation}\label{densityelements}
\begin{array}{l}
\rho_{11}(t)=\rho_{11}(0)|h_A(t)|^2|h_B(t)|^2,
\\
\rho_{22}(t)=|h_A(t)|^2[\rho_{22}(0)+\rho_{11}(0)(1-|h_B(t)|^2)],
\\
\rho_{33}(t)=|h_B(t)|^2[\rho_{33}(0)+\rho_{11}(0)(1-|h_A(t)|^2)],
\\
\rho_{44}(t)=1-(\rho_{11}(t)+\rho_{22}(t)+\rho_{33}(t)),
\end{array}
\end{equation}
and the nondiagonal elements are
\begin{equation}\label{densitynelements}
\begin{array}{l}
\rho_{12}(t)=\rho_{12}(0)|h_A(t)|^2h_B(t),
\\
\rho_{13}(t)=\rho_{13}(0)h_A(t)|h_B(t)|^2,
\\
\rho_{14}(t)=\rho_{14}(0)h_A(t)h_B(t),
\\
\rho_{23}(t)=\rho_{23}(0)h_A(t)h_B(t),
\\
\rho_{24}(t)=h_A(t)[\rho_{24}(0)+\rho_{13}(0)(1-|h_B(t)|^2)],
\\
\rho_{34}(t)=h_B(t)[\rho_{34}(0)+\rho_{12}(0)(1-|h_A(t)|^2)],
\end{array}
\end{equation}
and $\rho_{ij}(t)=\rho^{*}_{ji}(t)$. We should note that if we let
$h_A(t)=h(t)$ and $h_B(t)=1$, this reduced density matrix represents
the dynamics of the two qubits in the single-sided channel case.

\section{Entanglement evolution}\label{entdynamic}
In this section, we give the entanglement evolution for single-sided
channel case ($h_A(t)=h(t)$, $h_B(t)=1$) and two-sided channels case
for pure and mixed initial states. Firstly, we assume that the
qubits are initially prepared in the pure state
\begin{equation}\label{pureinitial}
  |\Psi\rangle=c_1|ee\rangle+c_2|eg\rangle+c_3|ge\rangle+c_4|gg\rangle,
\end{equation}
where $c_i(i=1,2,3,4)$ are complex and satisfy the normalization
condition $\sum^4_{i=1}|c_i|^2=1$.  The elements of the density
matrix corresponding to this initial state can be written as
$\rho_{ij}(0)=c_ic_j^*,(i,j=1, 2, 3, 4)$. In what follows, we
consider the single-sided channel case. Substituting $h_A(t)=h(t)$,
$h_B(t)=1$ and $\rho_{ij}(0)=c_ic_j^*$ into Eqs.
(\ref{densityelements}) and (\ref{densitynelements}) we obtain the
diagonal elements
\begin{equation}\label{spuredi}
\begin{array}{l}
\rho_{11}(t)=c_1^*c_1|h(t)|^2,
\\
\rho_{22}(t)=c_2^*c_2|h(t)|^2,
\\
\rho_{33}(t)=c_3^*c_3+c_1^*c_1(1-|h(t)|^2),
\\
\rho_{44}(t)=c_4^*c_4+c_2^*c_2(1-|h(t)|^2),
\end{array}
\end{equation}
and the nondiagonal elements
\begin{equation}\label{spurendi}
\begin{array}{l}
\rho_{12}(t)=c_1c_2^*|h(t)|^2,
\\
\rho_{13}(t)=c_1c_3^*h(t),
\\
\rho_{14}(t)=c_1c_4^*h(t),
\\
\rho_{23}(t)=c_2c_3^*h(t),
\\
\rho_{24}(t)=c_2c_4^*h(t),
\\
\rho_{34}(t)=c_3c_4^*+c_1c_2^*(1-|h(t)|^2),
\end{array}
\end{equation}
and $\rho_{ij}(t)=\rho^{*}_{ji}(t)$.

To quantify the entanglement we use Wootters
concurrence\cite{Wootters}. From Eqs. (\ref{spuredi}) and
(\ref{spurendi}), we find that the concurrence of the two qubits is
\begin{equation}\label{entanglementpure}
C_{\Psi}(t)=C_{\Psi}(0)|h(t)|,
\end{equation}
where $C_{\Psi}(0)=2|c_1c_4-c_2c_3|$ is the initial entanglement of
state (\ref{pureinitial}). Equation (\ref{entanglementpure}) shows
that the entanglement can be factorized as two terms whose physical
meaning is given as follows. We consider the extreme case in which
the two qubits are initially prepared in the maximally entangled
state, i.e., $C_{\Psi}(0)=1$. Then equation (\ref{entanglementpure})
reduces to $C_{\Psi}(t)=|h(t)|$, that means that $|h(t)|$ stands for
the time evolution of the entanglement for the maximally entangled
initial state under the single-sided channel. Then we can say that
the entanglement reduction under single-sided noisy channel is equal
to the product of the initial entanglement $C_{\Psi}(0)$ and the
time evolution of the entanglement for the maximally entangled
initial state. That is, the dynamics of the entanglement is
completely determined by the time evolution of the entanglement for
the initial state with maximal entanglement.

The factorization law given in Eq.(\ref{entanglementpure}) is valid
for single-sided channel case. One may ask to what extent it can be
generalized to the two-sided channels case. In what follows, we will
show that this can be done by constraining the initial state to a
certain class of states. We also let the two qubits be initially
prepared in the pure state (\ref{pureinitial}). Substituting
$\rho_{ij}(0)=c_ic_j^*$ into Eqs. (\ref{densityelements}) and
(\ref{densitynelements}) and after some calculations we find that
the concurrence at time $t$ is given by
\begin{equation}\label{entanglementpure2}
C_{\Psi}(t)=\textbf{max}\{0,Q(t)\}.
\end{equation}
The explicit expression of $Q(t)$ is
\begin{equation}\label{Q(t)}
Q(t)=C_{\Psi}(0)|h_A(t)||h_B(t)|\mathcal{X},
\end{equation}
where $\mathcal{X}=1-|c_1|^2/|c_2c_3-c_1c_4|\sqrt{\xi}$ with
$\xi=(1-|h_A(t)|^2)(1-|h_B(t)|^2)$.  Equation
(\ref{entanglementpure2}) shows that for the case of two-sided noisy
channels, the entanglement evolution is not equal to the product of
the initial entanglement $C_{\Psi}(0)$ and the maximal entanglement
evolution of single-sided channel $|h_A(t)|$ and $|h_B(t)|$ but an
additional term $\mathcal{X}$ should be taken into account. We
should note that if for some time intervals $\mathcal{X}<0$, the
entanglement sudden death occurs. Interestingly if we consider the
case $c_1=0$, that is, the two qubits are initially prepared in the
state
\begin{equation}\label{pureinitial2}
|\Phi\rangle=c_2|eg\rangle+c_3|ge\rangle+c_4|gg\rangle,
\end{equation}
with $\sum^4_{i=2}|c_i|^2=1$, we find that $\mathcal{X}=1$ and
Eq.(\ref{entanglementpure2}) reduces to
\begin{equation}\label{epurer}
C_{\Phi}(t)=C_{\Phi}(0)|h_A(t)||h_B(t)|,
\end{equation}
with $C_{\Phi}(0)=2|c_2c_3|$ being the entanglement of initial state
(\ref{pureinitial2}). Equation (\ref{epurer}) indicates that for the
pure initial state (\ref{pureinitial2}), the entanglement evolution
under two-sided noisy channels is equal to the product of the
initial entanglement $C_{\Phi}(0)$, $|h_A{t}|$ (the maximal
entanglement evolution under single-sided channel $A$) and
$|h_B{t}|$ (the maximal entanglement evolution under single-sided
channel $B$).

As we know, quantum mechanics and quantum information processing are
not constrained to pure states. So we give the generalization of the
factorization law to the mixed state. We consider the case in which
the qubits are initially prepared in the mixed state
\begin{equation}\label{mixedinitial}
\rho(0)=\left(
\begin{array}{cccc}
0&0&0&0\\
0&b&z&e\\
0&z^*&c&f\\
0&e^*&f^*&d\\
\end{array}
\right)\\
\end{equation}
where $b,c$ and $d$ are real numbers and satisfy $b+c+d=1$. For this
state it is easy to check that the initial concurrence
$C_{\rho}(0)=\textbf{max}\{0,\sqrt{bc}+|z|-|\sqrt{bc}-|z||\}$ and
the terms $\rho_{ij}(0)$ in Eqs.(\ref{densityelements}) and
(\ref{densitynelements}) can be written as
\begin{equation*}
\begin{array}{cccc}
\rho_{11}(0)=0,&\rho_{12}(0)=0,&\rho_{13}(0)=0,&\rho_{14}(0)=0,\\
\rho_{21}(0)=0,&\rho_{22}(0)=b,&\rho_{23}(0)=z,&\rho_{24}(0)=e,\\
\rho_{31}(0)=0,&\rho_{32}(0)=z^*,&\rho_{33}(0)=c,&\rho_{34}(0)=f,\\
\rho_{41}(0)=0,&\rho_{42}(0)=e^*,&\rho_{43}(0)=f^*&\rho_{44}(0)=d.\\
\end{array}
\end{equation*}
After some calculations we find that the concurrence at time $t$ is given by
\begin{equation}\label{entanglement2}
C_{\rho}(t)=C_{\rho}(0)|h_A(t)||h_B(t)|.
\end{equation}
That is the factorization law for the two-sided noisy channel case.
We should note that the initial states (\ref{pureinitial2}) and
(\ref{mixedinitial}) have a common property, that is, the number of
excitations in the system is not more than one. We denote these
states NOE states. Then we can conclude that for the two-sided noisy
channels case, when the qubits are initially prepared in NOE states,
no matter pure or mixed, the concurrence can be expressed as the
products of initial concurrence and the channels' action on the
maximally entangled state.

\section{Applications}\label{applications}
In this section we give two applications of our results. The first
one is the entanglement preservation. Equations
(\ref{entanglementpure2}) and (\ref{entanglement2}) show the
direction for entanglement preservation, that is, choosing
appropriate parameters to make the values of $|h_A(t)|$ and
$|h_B(t)|$ around 1. As an example, we investigate the detuning case
of a Lorentian spectral density\cite{Breuerb}
\begin{equation}\label{ff}
J(\omega)=\frac 1
{2\pi}\frac{\gamma_0\lambda^2}{(\omega_0-\omega-\Delta)^2+\lambda^2},
\end{equation}
where $\Delta=\omega_0-\omega_c$ is the detuning of $\omega_c$ and
$\omega_0$, and $\omega_c$ is the center frequency of the cavity.
$\lambda$ defines the spectral width of the reservoir and is
connected to the reservoir correlation time $\tau_R=\lambda^{-1}$.
$\gamma_0$ is related to the decay of the excited state of the qubit
in the Markovian limit of a flat spectrum, the relaxation time scale
$\tau_S$ over which the state of the system changes is then related
to $\gamma_0$ by $\tau_S=\gamma_0^{-1}$. We evaluate the reservoir
correlation function $f(t-t_1)$ using the spectral density
$J(\omega)$ and obtain
\begin{equation}\label{ff}
f(t-t_1)=\frac 1 2\gamma_0\lambda\exp[-(\lambda-i\Delta)(t-t_1)].
\end{equation}
Solving equation (\ref{difc1}) with this correlation function, we find
\begin{equation}\label{hexpression}
h(t)=e^{-\frac 1 2(\lambda-i\Delta)t}\biggr[\cosh\biggl(\frac{d t} 2 \biggr)
+\frac{\lambda-i\Delta}d\sinh\biggl(\frac{d t} 2\biggr)\biggr]
\end{equation}
with $d=\sqrt{(\lambda-i\Delta)^2-2\gamma_0\lambda}$. From
Eq.(\ref{hexpression}) we find that the value of $|h(t)|$ is
determined by the parameters $\Delta$ and $\lambda$. For the
two-sided channel, we denote $\Delta_A$ and $\lambda_A$ for channel
$A$ and $\Delta_B$ and $\lambda_B$ for channel $B$.

\begin{figure}
\includegraphics*[width=0.90\columnwidth,
height=0.85\columnwidth]{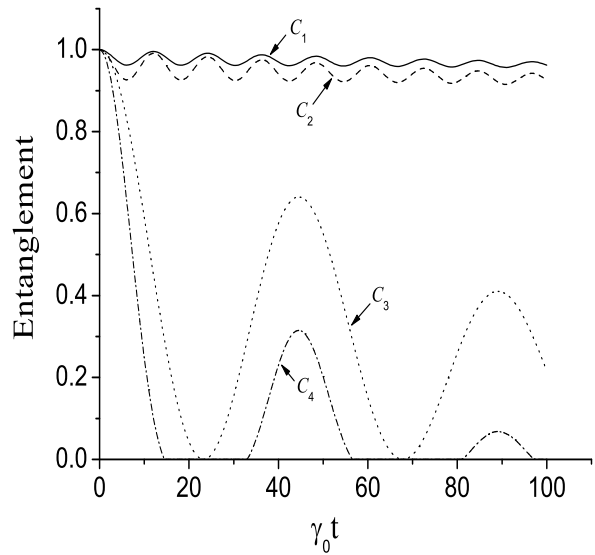} \vspace{-0.3cm} \caption{Dynamics
of entanglement as functions of $\gamma_{0}t$.}\label{fig1}
\vspace{-0.3cm}
\end{figure}

Figure \ref{fig1} shows the dynamics of the entanglement as
functions of $\gamma_{0}t$ for different cases. In Fig.\ref{fig1},
$C_1$ is the entanglement evolution of the initial state
$\psi=1/\sqrt 2(|eg\rangle+|ge\rangle)$ for the single-sided channel
case and $C_2$ is that for the two-sided channel case. We choose the
parameters $\Delta_A=\Delta_B=0.5\gamma_0$ and
$\lambda_A=\lambda_B=0.01\gamma_0$ to make the values of $|h_A(t)|$
and $|h_B(t)|$ around 1. Clearly, the entanglement is protected
well.

In Fig.\ref{fig1}, we also give the entanglement evolutions for
two-sided channel case of initial state $\psi=1/\sqrt
2(|eg\rangle+|ge\rangle)$ ($C_3$) and initial state $\phi=1/\sqrt
2(|ee\rangle+|gg\rangle)$ ($C_4$). Here, we choose the parameters
$\Delta_A=\Delta_B=0$ and $\lambda_A=\lambda_B=0.01\gamma_0$.
Comparing $C_3$ and $C_4$ we find that for initial state $\psi$ (NOE
sate), the entanglement is always large than zero except some
discrete time instants. This can be understood as follows. For the
NOE sates the factorization law (\ref{epurer}) is valid. The
entanglement is determined by $|h_A(t)|$ and $|h_B(t)|$ of which the
values are not less than zero for all the parameters. However, for
the initial state $\phi$ (not NOE sate), the entanglement sudden
death and the entanglement sudden birth can occur. This can be
understood from Eqs.(\ref{entanglementpure2}) and (\ref{Q(t)}) that
in this case $\mathcal{X}$ can be less than zero for some time
intervals.

The second application is to show the connection of the entanglement
and the non-Markovianity of the channels. The measure for
non-Markovianity defined in Ref. \cite{BreuerNd} is
\begin{equation}\label{non-Markovianity}
\mathcal{N}=\max_{\rho_{1,2}(0)}\int_{\sigma>0}dt\sigma(t),
\end{equation}
where $\sigma(t)$ is the rate of change of the trace distance which can be defined as
\begin{equation}
\sigma(t)=\frac d{dt}D\bigl(\rho_1(t),\rho_2(t)\bigr),
\end{equation}
with $D(\rho_1,\rho_2)=\frac12|\rho_1-\rho_2|$ being the trace
distance of the quantum states $\rho_1$ and $\rho_2$. Here,
$D(\rho_1,\rho_2)$ describes the distinguishability between the two
states and satisfying $0\leq D \leq1$. We should note that if there
exists a pair of initial states and a certain time $t$ such that
$\sigma(t)> 0$, the process is non-Markovian. Physically, this means
that for non-Markovian dynamics the distinguishability of the pair
of states increases at certain times. This can be interpreted as a
flow of information from the environment back to the system which
enhances the possibility of distinguishing the two states. Following
Ref.\cite{BreuerNd}, reference \cite{zyxu} presented a practical
idea for directly measuring the non-Markovian character of a single
qubit coupled to a zero-temperature bosonic reservoir. It is easy to
check that any pair of initial states satisfying the conditions in
the theorem given in Ref.\cite{zyxu} definitely owns the same
maximal trace distance
\begin{equation}\label{trace distance}
D\bigl(\rho_1(t),\rho_2(t)\bigr)=|h(t)|,
\end{equation}
namely, $|h(t)|$ stands for the maximal trace distance of one part
($A$ or $B$) of our model. From Eq.(\ref{entanglementpure}) we can
find that the entanglement evolution under single-sided noisy
channel is equal to the product of the initial entanglment
$C_{\Psi}(0)$ and the maximal trace distance $|h(t)|$. We also know
that $|h(t)|$ also stands for the channels' action on the maximally
entangled state. It is evident from this relation that any increase
of the entanglement implies occurring of the non-Markovian dynamics.
Hence a conceptually simple way to quantify the degree of
non-Markovianity of an unknown quantum evolution would be to compute
the amount of entanglement at different times within a selected
interval and check for strict monotonic decrease of the quantum
correlations. This measurement has been given in the very recent
paper\cite{Rivas}.

\vspace{0.2cm}

\section{Conclusion}\label{conclusion}
In summary, we have studied the entanglement dynamics of two qubits
under local, single- and two-sided noisy channels. We find that in
the case where the initial state is pure, and only one of the
subsystems undergoes a non-Markovian channel, the equation of motion
for entanglement presents the form of a simple factorization law
--the second term contains only information about how the maximal
entanglement is affected by the dynamics, and the first term scales
the second by the initial amount of entanglement. For more realistic
scenarios, where both parts are influenced by the local
environments, when the qubits are initially prepared in NOE states,
no matter pure or mixed, the concurrence can be expressed as the
products of initial concurrence and the channels' action on the
maximally entangled state. Using these factorization laws we have
found the way to protect entanglement from decay and given a
connection of the entanglement and the non-Markovianity of the
channels.

\vspace{-0.2cm}
\begin{acknowledgments}
\vspace{-0.2cm} This work is financially supported by Fundamental
Research Fund of Beijing Institute of Technology (Grant No.
20081742010) and National Science Foundation of China (Grant
No.10974016, No.11005008 and No.11075013). J.G.L. wishes to thank
Xiu-San Xing for valuable discussions.
\end{acknowledgments}

\bibliography{mcwf}

\begin{thebibliography}{99}
\bibitem{Nielsen} M.A. Nielsen and I.L. Chuang, \textit{Quantum Computation and Quantum
Information}, (Cambridge University Press, Cambridge, UK, 2000).
\bibitem{Bennett}C.H. Bennett, D.P. DiVincenzo, Nature \textbf{404},
247(2000).
\bibitem{Breuerb}H.-P. Breuer and F. Petruccione, \textit{The Theory of Open
Quantum Systems},(Oxford University Press, Oxford, UK, 2002).
\bibitem{Gardinerb}C.W. Gardiner and P.
Zoller, \textit{Quantum Noise}, (Springer-Verlag, Berlin, Germany, 1999).
\bibitem{TYu} T. Yu and J. H. Eberly, Phys. Rev. Lett. \textbf{97}, 140403 (2006).
\bibitem{Dodd} P. J. Dodd and J. J. Halliwell, Phys. Rev. A \textbf{69}, 052105 (2004).
\bibitem{Zczkowski} K. \`{Z}yczkowski, P. Horodecki, M. Horodecki, and R. Horodecki,
Phys. Rev. A \textbf{65}, 012101 (2001).
\bibitem{Konrad} T. Konrad, F. de Melo, M. Tiersch, C. Kasztelan, A. Arag\~{a}o, and
A. Buchleitner, Nat. Phys. \textbf{4}, 99 (2008).
\bibitem{Chang} C.-S. Yu, X.-X. Yi, and H.-S. Song, Phys. Rev. A \textbf{78},
062330 (2008).
\bibitem{Zong} Z.-G. Li, S.-M. Fei, Z.-D. Wang, and W.-M. Liu, Phys. Rev. A
\textbf{79}, 024303 (2009).
\bibitem{Piilo1}J. Piilo, S. Maniscalco, K. H\"{a}rk\"{o}nen, and K.-A.
Suominen, Phys. Rev. Lett. \textbf{100}, 180402 (2008).
\bibitem{Breuer3}H.P. Breuer and J. Piilo, Europhys.
Lett. \textbf{85}, 50004 (2009).

\bibitem{Maniscalco} S. Maniscalco and F. Petruccione, Phys. Rev. A
\textbf{73}, 012111 (2006).
\bibitem{Ferraro}E. Ferraro, H.-P. Breuer,A. Napoli, M. A. Jivulescu, and
A. Messina, Phys. Rev. B \textbf{78}, 064309 (2008).
\bibitem{Budini} A. A. Budini, Phys. Rev. A \textbf{74}, 053815 (2006).

\bibitem{Dijkstra}A. G. Dijkstra and Y. Tanimura, Phys. Rev. Lett. \textbf{104}, 250401 (2010).

\bibitem{Vacchini}B. Vacchini,  Phys. Rev. A \textbf{78}, 022112 (2008).

\bibitem{Daffer}S. Daffer, K. W\'{o}dkiewicz, J.D. Cresser, and J.K. McIver, Phys. Rev.
A \textbf{70}, 010304 (2004).
\bibitem{Chou}C.-H. Chou, T. Yu, and B. L. Hu, Phys. Rev. E, \textbf{77}, 011112 (2008).
\bibitem{Huang}X.-L. Huang , H.-Y. Sun, and X.-X. Yi, Phys. Rev. E \textbf{78}, 041107
(2008).
\bibitem{Shabani}A. Shabani and D. A. Lidar, Phys. Rev. A \textbf{71}, 020101(R)
(2005).
\bibitem{Chru}D. Chru\'{s}ci\'{n}ski and A. Kossakowski, Phys.
Rev. Lett. \textbf{104}, 070406 (2010).
\bibitem{Khalili}F. Khalili, S. Danilishin, H.-X. Miao, H. M\"{u}ller-Ebhardt, H. Yang, and
Y.-B. Chen, Phys. Rev. Lett. \textbf{105}, 070403 (2010).
\bibitem{Fanchini} F. F. Fanchini, T. Werlang, C. A. Brasil, L. G. E. Arruda, and
A. O. Caldeira, Phys. Rev. A \textbf{81}, 052107 (2010).
\bibitem{Bellomo} B. Bellomo, R. Lo Franco, and G. Compagno, Phys.
Rev. Lett. \textbf{99}, 160502 (2007).
\bibitem{Wolf} M.M. Wolf, J. Eisert, T.S. Cubitt and J.I. Cirac, Phys. Rev. Lett, \textbf{101}, 150402
(2008).
\bibitem{BreuerNd} H.-P. Breuer, E.-M Laine and J. Piilo,
Phys. Rev. Lett. \textbf{103}, 210401 (2009).
\bibitem{Rivas} \'{A}. Rivas, S.F. Huelga and M.B. Plenio, Phys. Rev. Lett, \textbf{105}, 050403
(2010).
\bibitem{Laine}E.-M Laine,  J. Piilo and H.-P. Breuer,
Phys. Rev. A \textbf{81}, 062115 (2010).
\bibitem{Wootters} W.K. Wootters, Phys. Rev. Lett. \textbf{80}, 2245
(1998).
\bibitem{zyxu} Z.-Y. Xu, W.-L. Yang and M. Feng, Phys. Rev. A \textbf{81}, 044105
(2010).

\end{thebibliography}

\end{document}